# Dynamic and stochastic models of the evolution of aftershocks

A.V. Guglielmi

*Schmidt Institute of Physics of the Earth, Russian Academy of Sciences, Moscow, Russia*

**Abstract**

This paper is devoted to the theory of aftershocks. The history of discovery of the Omori law is briefly described, the initial formulation of the law is given in the form of an algebraic formula describing the decrease in the frequency of aftershocks over time. An important generalization of the Oiori formula which is widely used in modern seismology is presented. The generalized law is also expressed by an algebraic formula, but it contains an additional parameter, which makes it possible to more flexibly approximate the observational data. The alternative approach to the theoretical description of aftershocks is to use the differential evolution models. To simulate the averaged dynamics of aftershocks, it is proposed the Verhulst differential equation, also known as the logistic equation. It is shown that the decrease in the frequency of aftershocks with time at the first stage of evolution occurs according to the Omori hyperbola, i.e. in accordance with the original wording of the law. The paper also proposes a stochastic generalization of the equation for the evolution of aftershocks. A random function of time is added to the right-hand side of the logistic equation to simulate noise that affects a dynamic system. As a result, a stochastic Langevin equation was obtained, which can be used to simulate fluctuations in the frequency of aftershocks.

## Introduction

The repeated tremors called aftershocks occur in the epicentral zone after a strong earthquake [Kasahara, 1981]. For the first time the law of evolution of the aftershocks was formulated by Fusakichi Omori. This outstanding event took place 127 years ago [Omori, 1984]. It is appropriate to tell a little about the history of this discovery [Davison, 1924, 1930; Utsu, Ogata, Matsu'ura, 1995; Guglielmi, 2017].

In 1880, John Milne, one of the founders of modern seismology, has created the horizontal pendulum seismograph, an easy-to-use and sensitive enough device for recording earthquakes. While Milne worked in Tokyo at the invitation of the Government of the Japanese Empire. In 1887 Milne was elected a Fellow of the Royal Society of London. He manages to convince the Royal Society to allocate funds for the creation of a global network of seismic stations equipped with his instruments. (By the way, the three stations Milne decided to place in Russia.) Fusakichi Omori was a student of John Milne. He enjoyed the encouraging support of his teacher, like all the young Japanese seismologists of the time. An earthquake with magnitude of M = 8 occurred on October 28, 1891. Milne's seismographs have recorded numerous aftershocks. The analysis of aftershocks allowed Omori to formulate the law that bears his name [Omori, 1984]. He was then 26 years old.

At the end of a small excursion into history, it is worth mentioning that Japan appreciated the merits Milne before the country and the world. Emperor Meiji awarded him the Order of the Rising Sun and awarded him a life pension, and the University of Tokyo elected him the honorary professor.

Omori law states that after a strong earthquake the frequency of aftershocks

$$n(t) = \frac{k}{c+t} \qquad (1)$$

on average hyperbolically decreases with time. Here the parameter $k > 0$ characterizes the event, $t \geq 0$, and the parameter $c > 0$ being determined by the choice of the initial condition $n(0)$ [Omori, 1984]. 127 years after the discovery of the law, the modeling of the evolution of aftershocks has already become a fairly established field, in which a large volume and rich in interesting content observational material has been accumulated which generally confirms the Omori law.

An important stage in the development of this area began after the publications [Utsu, 1957, 1961] (see also reviews [Utsu, Ogata, Matsu'ura, 1995; Guglielmi, 2017]). Utsu showed that the formula

$$n(t) = \frac{k}{(c+t)^p}. \qquad (2)$$

better approximates observational data than formula (1). Here $p$ is some fitting parameter. It should be noted that formula (2) was considered earlier in papers [Hirano, 1924; Jeffreys, 1938], but only Utsu fully appreciated its capabilities for flexible modeling of the evolution of aftershocks.

Formula (2) is widely used in the quantitative description of aftershocks. To date, extensive information has been accumulated on the $k$ and $p$ parameters for specific events that have occurred in many seismically active regions of the planet. Here is a far from complete list of relevant works [Adams, Le Fort, 1963; Page, 1968; Usami, 1975; McGarr, Green, 1978; Iio, 1984; Ogata, 1994; Kocharyan, Spivak, 2003; Kanamori, Brodsky, 2004; Rodkin, 2008; Narteau et al. 2009; Baranov, Chebrov, 2012; Romashkova, Kosobokov, 2012; Saltykov, Droznina, 2014; Sobolev, 2015; Goda, 2015; Zotov, Zavyalov., Klain, 2020]. Of particular interest is the variability of the parameter $p$. It varies from case to case, but the average over the set of many measurements is quite close to unity.

In this paper, we try to find an another approach to the problem of the evolution of aftershocks. Namely, instead of choosing an algebraic formula for approximating observations, we will try to find simple differential equations describing evolution. For example, it is quite obvious that the Omori's law (1) is equivalent to the truncated Bernoulli equation

$$\dot{n} + \sigma n^2 = 0. \qquad (3)$$

Here, the dot over the symbol denotes time differentiation, and $\sigma = 1/k$ is the so-called deactivation coefficient of the earthquake source [Guglitlmi, 2016]. The advantage of the simple differential model (3) compared with the algebraic formula (1) consists in the fact that the deactivation coefficient $\sigma(t)$ can be considered time-dependent in order to take into account the non-stationary of rocks in the source, "cooling down" after the main earthquake shock. In this case, the solution of equation (3) takes the form

$$n(t) = n_0 \left[ 1 + n_0 \int_0^t \sigma(t') dt' \right]^{-1}. \qquad (4)$$

Note that formula (1) is a special case of expression (4) and can be obtained from it at $\sigma = \text{const}$.

The formulation of Omori's law in the form of a differential equation suggests possible generalizations of the theory. Two generalizations are proposed below. One of them has the form of the logistic Verhulst equation, and the second formally coincides with the stochastic Langevin equation. The logistic equation models the averaged dynamics of aftershocks. Stochastic generalization is of interest in describing not only averaged evolution, but also fluctuations of aftershocks.

**Evolution equation**

In the paper [Faraoni, 2020] attention is drawn to the fact that equation (3) can be considered as the Euler-Lagrange equation of some dynamic system. The corresponding Lagrangian has the form

$$L(n, \dot{n}) = n\dot{n}^2 + \sigma^2 n^5. \qquad (5)$$

As is known, the Lagrange theory is based on the fundamental laws of Newtonian mechanics. In many specific tasks these laws allow us to choose one or another form of the Lagrangian. For obvious reason it is practically impossible to find the Lagrangian based on first principles in our case. The theory of aftershocks is purely phenomenological [Guglielmi, Klain, 2020]. Therefore, if we try to formulate it on the basis of the Euler-Lagrange equation, then the form of the Lagrangian must be guessed guided by observation data and indirect considerations.

Let us focus on the development of the theory of aftershocks in the direction indicated by Valerio Faraoni. Let us pay attention to the fact that Lagrangian (5), from which (3) follows, has an unusual structure. This structure suggests that $L(n,\dot{n})$ is represented as a series in odd powers of $n$. When searching for a more general theory is natural to try first of all to restore the cubic term of the form $\gamma^2 n^3$, which is missing in (5). After some trial and error the following Lagrangian was chosen:

$$L(n,\dot{n}) = n\left[\dot{n}^2 + n^2(\gamma - \sigma n)^2\right]. \qquad (6)$$

It is seen that (6) transforms into the Faraoni Lagrangian (5) at $\gamma = 0$.

The Euler-Lagrange equation

$$\frac{\partial L}{\partial n} - \frac{d}{dt}\frac{\partial L}{\partial \dot{n}} = 0 \qquad (7)$$

with the new Lagrangian (6) becomes

$$\dot{n} = n(\gamma - \sigma n). \qquad (8)$$

We got the Verhulst equation, or, as it is also called, the logistic equation [Verhulst, 1838]. It is widely used in biology, chemistry, astrophysics, and other sciences, including economics and sociology. Let us show that the logistic equation can also be useful in the physics of earthquakes.

The solution of equation (8) is the so-called logistic curve. It consists of two branches. Recall the form of the branch of the logistic curve, which is widely used in various sections of science. This is a monotonically increasing function of time. When $t \to -\infty$ function asymptotically approach zero, and when $t \to +\infty$ it seeks from the bottom to saturation at the $n_\infty = \gamma/\sigma$ level. This behavior is not characteristic of aftershoks.

The second branch, the existence of which is not so widely known, is entirely located above the saturation level and is a monotonically decreasing function of time. When $t \to +\infty$ it tends asymptotically from above the saturation level. It is this branch that is of interest for the theory of aftershocks.

We pose the Cauchy problem for the logistic equation. The selection of one or another branch is carried out by choosing the initial condition. We will set the initial condition $n = n_0$ at $t = 0$ and will look for solutions at $t > 0$. It is easy to show that for $n_0 < n_\infty$ ($n_\infty < n_0$) solution of the problem will be lower (upper) branch of the logistic curve. Thus, when setting the Cauchy problem in the physics of aftershocks, the initial condition should be asked the additional restriction $n_0 > n_\infty$. Moreover, it is quite appropriate to use a strong inequality

$$n_0 \gg n_\infty = \gamma/\sigma. \tag{9}$$

Indeed, when $t \to \infty$ the frequency of aftershocks asymptotically approaches to the background (equilibrium) value $n_\infty$. Experience shows that as a rule after a strong earthquake $n_0 \gg n_\infty$.

Let us show that the decrease in the frequency of aftershocks with time, which generally proceeds along the upper branch of the logistic function, at the first stage of evolution occurs according to the Omori hyperbola, i.e. in accordance with formula (4) at $\sigma = \text{const}$. It is natural to call this stage of evolution the Omori epoch.

Let us assume for simplicity that $\gamma = \text{const}$, $n_\infty = \text{const}$, introduce the notation

$$t_\infty = \frac{1}{\gamma} \ln\left(1 - \frac{n_\infty}{n_0}\right), \tag{10}$$

and write the solution of evolution equation (8) in the following form:

$$n(t) = n_\infty \left\{1 - \exp\left[\gamma(t_\infty - t)\right]\right\}^{-1}. \tag{11}$$

In the Omori epoch $t_\infty < t \ll 1/\gamma$ and respectively

$$n(t) = 1/\sigma(t - t_\infty). \tag{12}$$

Formula (12) coincides with the classical Omori formula (1) up to notation.

Observational experience indirectly testifies to the plausibility of our logistic model. It is known for example that over time the frequency $n$ tends not to zero, as follows from the Omori law, but to some equilibrium value $n_\infty$.

We presented an idealized picture of the evolution of aftershocks. In reality the frequency of excitation of aftershocks is influenced by many factors that are difficult to control. We mentioned above the nonstationarity of the geological environment in the source. Among other causes of deviation from the ideal picture it is necessary to highlight the impact of endogenous and exogenous triggers on the earthquake source. These include free oscillations of the Earth excited by the main shock [Guglielmi, Zotov, 2012; Guglielmi, 2015a], round-the-world seismic echo [Guglielmi, 2015b], seismic noises in the source, and others [Zavyalov, Guglielmi, Zotov, 2020].

## Discussion

We want to discuss the issue of a stochastic generalization of the evolution equation. Let's change the variable $n(t) \to \varphi(t) = 1/n(t)$, rewrite equation (8) in the form

$$\dot{\varphi} = \sigma - \gamma\varphi, \tag{13}$$

and add the function $\xi(t)$ to the right-hand side:

$$\dot{\varphi} = \sigma - \gamma\varphi + \xi(t). \tag{14}$$

Let us formally solve this equation

$$\varphi(t) = \varphi_\infty + (\varphi_0 - \varphi_\infty)\exp(-\gamma t) + \int_0^t \xi(t_1)\exp[\gamma(t_1 - t)]dt_1. \tag{15}$$

Here $\varphi_\infty = \sigma/\gamma$, $\varphi_0 = \varphi(0)$. We will consider $\xi(t)$ the random delta-correlated function with zero average value, i.e. we put

$$\overline{\xi(t)} = 0, \ \overline{\xi(t)\xi(t_1)} = 2D\delta(t - t_1), \tag{16}$$

where the line from above means averaging. Now equation (14) should be considered as the Langevin stochastic equation [Klimontovich, 1982]. The new phenomenological parameter $D$ is determined by the intensity of noise affecting our dynamic system.

So, the relaxation of the earthquake source after the main shock is manifested in a sequence of aftershocks, the frequency of which decreases with time. Observational experience shows that a particular sequence is characterized by a unique combination of numerous properties. One of the important tasks of the theory is to compact display the variety of properties in the form of a small set of phenomenological parameters. We see that the logistic evolution model contains two parameters, $\sigma$ and $\gamma$. In the Langevin model the third parameter $D$ is added to this set. Let us consider how these parameters can be estimated practically from the observation of aftershocks.

The most reliable is the deactivation coefficient in the Omori epoch: $\sigma = d\overline{\varphi}/dt$. Considerable experience has already been accumulated in measuring $\sigma$, and some connection between $\sigma$ and the magnitude of the main earthquake shock was noticed (see [Zavyalov, Guglielmi, Zotov, 2020] and the literature cited therein). A rough estimate of the parameter $\gamma$ can be made using the formula $\gamma \sim \sigma/\overline{\varphi_\infty}$. As for the $D$ parameter, the procedure for its measurement has yet to be developed. The general idea is that $D$ can be expressed in terms of $\gamma$ and the dispersion of the aftershock frequency (see, for example, [Klimontovich, 1982]).

Experimental study of variations in phenomenological parameters can enrich the physics of aftershocks. The variation of the deactivation coefficient is of particular interest. In the Omori epoch, the $d\overline{\varphi}/dt$ is constant by definition. From a theoretical point of view, at the end of the Omori epoch, the $d\overline{\varphi}/dt$ value monotonically decreases and tends to zero at $t \to \infty$. Experience however shows that sometimes the value $d\bar{\varphi}/dt = 0$ is reached at $t = t_c$, i.e. over a finite time interval [Zavyalov, Guglielmi, Zotov, 2020]. Analysis of the evolution equation indicates a phase transition of the dynamical system at $t = t_c$ [Guglielmi, Klain, 2020]. As $t$ approaches $t_c$, anomalous growth of seismicity fluctuations, critical deceleration (the so-called "mode softening"), and other interesting phenomena related to critical opalescence are expected [Guglielmi, 2015c].

**Conclusion**

This paper presented two models of the evolution of aftershocks – the logistic model which describes the variation in the frequency of aftershocks, and the Langevin model, which takes account the fluctuations of aftershocks. A characteristic feature of the models is that they are differential. The models contain a small set of phenomenological parameters that quantitatively characterize each specific sequence of aftershocks. Models parameters can, in principle, be measured experimentally.

**Acknowledgments**. I express my deep gratitude to B.I. Klain for the joint development of theoretical aspects of the problem, as well as A.D. Zavyalov and O.D. Zotov, together with whom a series of experimental studies on modeling the dynamics of aftershocks was carried out. The work was done within the framework of the state assignments programs of the IPhE RAS.

*Information about the author*

**GUGLIELMI Anatol Vladimirovich** – Dr. Sci. (Phys.-Math.), Prof., Chief Researcher, Institute of Physics of the Earth, RAS, 10 B. Gruzinskaya 123995, Moscow, Russia, e-mail: guglielmi@mail.ru.